\documentclass[12pt,letterpaper]{article}
\usepackage{amsfonts}
\usepackage{amsmath}
\usepackage{amssymb}
\usepackage{color}
\usepackage{graphicx}
\usepackage{young}
\usepackage[vcentermath]{youngtab}
\usepackage{stackengine}
\usepackage{bbm}
\usepackage{diagbox}
\usepackage{slashed}
\usepackage{subcaption}
\usepackage[utf8]{inputenc}
\usepackage{graphicx}
\graphicspath{ {./images/} }
\usepackage{bbm}

\usepackage{tabularx}

\usepackage[colorinlistoftodos]{todonotes}

\usepackage{lipsum}

\usepackage{hyperref}
\usepackage[left=2.5cm,right=2.5cm,top=2.5cm,bottom=3cm]{geometry}
\numberwithin{equation}{section}

\def\bea#1\eea{\begin{align}#1\end{align}}
\def\bes #1\ees{\begin{split}#1\end{split}}
\newcommand{\be}{\begin{equation}}
\newcommand{\ee}{\end{equation}}

\usepackage[sorting=none]{biblatex}
\usepackage{epigraph}

\usepackage{etoolbox}

\makeatletter
\patchcmd{\epigraph}{\@epitext{#1}}{\itshape\@epitext{#1}}{}{}
\makeatother

\begin{document}

% \title{The $N$-ary in the Coal Mine: Avoiding Mixture Model Failure with Proper Validation}
% \maketitle

\begin{titlepage}

\begin{center}

\today
\hfill         
\phantom{xxx}  Preprint-\#\#\#

\vskip 2 cm {\Large \bf The $N$-ary in the Coal Mine: Avoiding Mixture Model Failure with Proper Validation}

\vskip 1.25 cm {\bf Travis Maxfield$^{a}$, Joshua Hochuli$^{a}$, James Wellnitz$^{a}$, Cleber Melo-Filho$^{a}$, Konstantin I. Popov$^{a}$, Eugene Muratov$^{a,b,\ast}$, and Alex Tropsha$^{a,\ast}$}\\

\vskip 0.2 cm
%$^{a}${\it Department of Electrical and Computer Engineering, Duke University, Durham, NC, USA}
%\vskip 0.2 cm
$^{a}${\it UNC Eshelman School of Pharmacy, University of North Carolina, Chapel Hill, NC, USA}
\vskip 0.2 cm
$^{b}${\it Federal University of Paraiba, Joao Pessoa, PB, Brazil}

\vskip 0.2 cm
{$^\ast$ Corresponding Authors:}
\href{mailto:alex_tropsha@unc.edu}{alex\_tropsha@unc.edu}, 
\href{mailto:murik@email.unc.edu}{murik@email.unc.edu}

\end{center}
\vskip 1 cm

\begin{abstract}
\baselineskip=18pt
Modeling the properties of chemical mixtures is a difficult but important part of any modeling process intended to be applicable to the often messy and impure phenomena of everyday life, including food and environmental safety, healthcare, etc. Part of this difficulty stems from the increased complexity of designing suitable model validation schemes for mixture data, a fact which has been elucidated in previous work only in the case of binary mixture models. We extend these previously defined validation strategies for QSAR modeling of binary mixtures to the more complex case of general, $N$-ary mixtures and argue that these strategies are applicable to many modeling tasks beyond simple chemical mixtures. Additionally, we propose a method of establishing a baseline model performance for each mixture dataset to be in used in model selection comparisons. This baseline is intended to account for the statistical dependence generically present between the properties of mixtures that share constituents. We contend that without such a baseline, estimates of model performance can be dramatically overestimated, and we demonstrate this with multiple case studies using real and simulated data.

\end{abstract}

\end{titlepage}

\tableofcontents

\section{Introduction: The Problem of Mixture Modeling}

\epigraph{``The first principle is that you must not fool yourself---and you are the easiest person to fool."}{--- \textup{Richard Feynman}}

Chemical mixtures are ubiquitous in natural and industrial settings. Therefore, it often behooves modelers to incorporate into their modeling the assessment of those properties inherent to a mixture, as distinct from the properties of the individual chemical constituents. In other words, the modeling of chemical mixtures is a crucial component of chemical safety determination, evaluation of drug-drug interactions, design of drug delivery systems, and development of combination therapies for complex diseases. Furthermore, as pervasive as chemical mixtures are in everyday phenomena, the concept of a mixture is more so. Myriad applications in cheminformatics and bioinformatics modeling can be cast as essentially problems in mixture modeling. These include modeling concentration-dependent properties of general chemical mixtures, protein-ligand interactions, and chemical fragment-based modeling, among others.

It is therefore a meaningful challenge to develop approaches to provide accurate predictions of mixture properties from their composition. Early efforts in this direction include~\cite{lu_joint_2009,lin_quantification_2003,altenburger_mixture_2003,chatterjee_prediction_2021} as well as work by a subset of the current authors on the development of descriptors for chemical mixtures~\cite{muratov_existing_2012}. These approaches were applied to modeling of mixtures of organic solvents~\cite{muratov_existing_2012}, drug delivery systems~\cite{alves_cheminformatics-driven_2019}, inorganic materials~\cite{isayev_universal_2017}, and drug-drug interactions~\cite{zakharov_qsar_2016}, as well as the prediction of synergistic effects in drug mixtures generally~\cite{muratov_qsar_2013} and specifically as applied to SARS-CoV-2~\cite{bobrowski_synergistic_2021}.

However, part of any modeling task is model validation, which is complicated in the case of mixture data by the underappreciated fact that different mixtures sharing some but not all constituents tend to have correlated properties. We can make a genetics analogy and refer to this fact as `mixture heritability'. Without special care, mixture heritability can unpredictably affect model validation results. Model validation strategies that take heritability into account were proposed for binary mixtures in~\cite{muratov_jsm_2014,muratov_existing_2012}. These strategies amount to stratifying the validation datasets according to the number of shared constituents---or, to continue the genetics analogy, degree of relatedness---with mixtures in the model's training data.

Increasing access to mixture data---not only for binary but $N$-ary mixtures as well---has led to a growth of interest in the computational modeling of mixtures. Unfortunately, the validation strategies for binary mixture QSAR models previously described are not suitable for more complex, $N$-ary mixtures, as in~\cite{wang_comparative_2018} for example. Building on these validation strategies, we propose a generalization applicable broadly to $N$-ary mixtures. Additionally, we describe a novel performance assessment akin to the use of random pseudodescriptors but specific to mixture modeling~\cite{rucker_y-randomization_2007}. This performance assessment is meant to demonstrate the degree to which a model's performance relies on mixture heritability rather than structural-property relationships.

It is our contention that proper, rational validation of mixture models is not only important for academic model development but is crucial to any real-world application of a model, where a realistic notion of the model's performance is integral. While occasional model failure---by which we mean a significant discrepancy between expected and actual performance---is inevitable, some failures can and should be prevented. Failure due to improper validation is of the latter type, and so we hope that the validation strategies laid out in this and past papers will be like the alluded-to, titular canary was to coal miners: as crucial as pickaxes.

\section{Mixture Model Validation}\label{sec:validation}
Mixture model validation---inclusive of both the composition of training and validation datasets as well as those metrics used to assess the performance of the model on these sets---is heavily influenced by the purpose of the mixture model, by which we mean the intended number of shared constituents among mixtures in the internal and external data. For example, a modeler may be interested in finding a new drug to replace a single constituent of a mixture, and therefore the internal and external data will be designed to share all but one constituent. Alternatively, a modeler may desire an entirely new mixture composed of all novel constituents; this scenario suggests the internal and external data share no constituents. We will frequently refer to various intended uses for a model, and these examples (and their generalization) will be what we have in mind. 

Rational validation is designed to mimic a model's intended purpose, and so we must choose our validation sets to have the appropriate constituent overlap, or degree of relatedness, as we have previously referred to this concept. We will shortly demonstrate how this can be done in a simple example in which our dataset is composed of mixtures drawn from a single collection of drugs. Appendix~\ref{sec:mixtures} contains the full details of our proposal in generality. First, let us recall how validation is commonly performed on single-chemical datasets.

\subsection{Standard model validation}

The goal of any model validation procedure is to procure an estimate of the performance of the model on as-yet-unseen external data. Standard external validation of models built on single-chemical datasets proceeds usually by $k$-fold cross-validation: splitting the dataset randomly into $k$ disjoint subsets\footnote{$k$ is often taken equal to $5$.}, called `folds', training a model on the union of $k-1$ folds and assessing the model's performance on the held-out fold. This is repeated $k$ times with a different held-out fold each time, producing in the end $k$ model scores---accuracy, positive predictive value, etc. Each score is presumed to be indicative of the true performance of the model on external data, but this requires certain fundamental assumptions about the training, validation, and external data.

Most germane to this discussion is the assumption that the validation data---each of the held-out folds in $k$-fold cross-validation---is sampled from the same statistical distribution as the external data. One way that this assumption can fail, which is already familiar to QSAR modelers, is if the external data is very dissimilar to the validation data. The incorporation of applicability domains into QSAR models~\cite{weaver_importance_2008} can mitigate this risk by alerting the model user to the presence of external data that is outside the scope of the model's training/validation distribution.

Another way that this assumption can fail, and the subject of this paper, is when training and validation data are statistically dependent or correlated in a way that the training and external data are not. This occurs generically for mixture data when standard validation is performed, because the properties of any given mixture will generically be correlated with those of mixtures containing a subset of its constituents. That is, mixtures `inherit' some of their properties from their constituents and therefore related mixtures---those sharing some but not all constituents---will typically have correlated properties. This is not always true, of course, but it is the responsibility of the modeler to account for the possibility and demonstrate its absence when applicable. We will provide prescriptions for doing so, both in terms of how to choose validation datasets as well as how to quantitatively estimate mixture heritability, in the following.

One wrinkle in the story so far is that sometimes external data is in fact correlated with the training and validation data by design. This is the case when a model's intended use is to predict properties only among related mixtures. In this case, the model learning heritability is not a bug but an essential feature. These cases can often be stratified by the degree of intended relatedness among the mixtures in the training/validation sets and the external set. By degree here, we refer to the number of shared constituents in the mixture. More shared constituents will lead to a higher correlation between the training and validation datasets generically. With this in mind, our proposed validation strategies are also stratified by the degree of correlation, extending the `compounds out' and `everything out' validation strategies applicable to binary mixtures~\cite{muratov_existing_2012,muratov_jsm_2014}. These strategies were applicable for external sets sharing a single mixture constituent and no mixture constituents, respectively.

\subsection{Training and validation datasets for mixture modeling}

Standard single-chemical validation, because it decomposes a dataset randomly into training and validation subsets, will almost certainly place related mixtures into different subsets when applied without modification to mixture data. When this occurs, a model can perform deceptively well on the validation data by learning the correlation or heritability among related mixtures instead of the underlying structural-property relationship. When faced with external, uncorrelated data, the model will fail without this crutch. Examples of such naive validation strategies abound~\cite{qin_qsar_2018,pan_prediction_2019,luan_prediction_2013,menden_community_2019,moesser_protein-ligand_2022}. Although, in some cases, the authors were aware of the flaws of using this approach, stating “it was not a validation, as potentially some of these compounds were included into the training set”~\cite{chushak_silico_2020}. However, they did not try to address this issue.

In its simplest setting---chemical mixtures whose constituents are all drawn from a single collection of drugs---the method we propose is very similar to $k$-fold cross-validation; however, instead of acting on the entire mixture dataset, we essentially perform the $k$-fold split on the collection of drug constituents, the precise meaning of which we detail presently. That is, for a single fold, we will split the collection of constituents into two disjoint subsets, which we dub the `interior' and `exterior' sets. Note that we do not use the terms internal and external here, as those terms are reserved for data, i.e.\ mixtures, and not for mixture constituents such as drugs in this example. Training is then performed on all mixtures composed of drugs in the interior set, and we define various forms of validation sets by collecting mixtures with differing numbers of constituents drawn from each of the interior and exterior sets.

To be even more specific, we will focus first on the binary mixture example and see how this language will reproduce the results of~\cite{muratov_existing_2012, muratov_jsm_2014}. In this case, there are two validation sets and a single training set. The latter is used for both validations consisting of all mixtures built from drugs in the interior set. The first validation set, termed `compounds out' in previous works, is composed of mixtures with a single constituent in the interior set of drugs and a single constituent in the exterior set. We call this `1 compound out'. The second, termed `everything out' previously, is composed of mixtures built entirely from the exterior set of drugs. We call this `2 compounds out'.

\begin{figure}[hbt!]
 \centering
 \includegraphics[width = \textwidth]{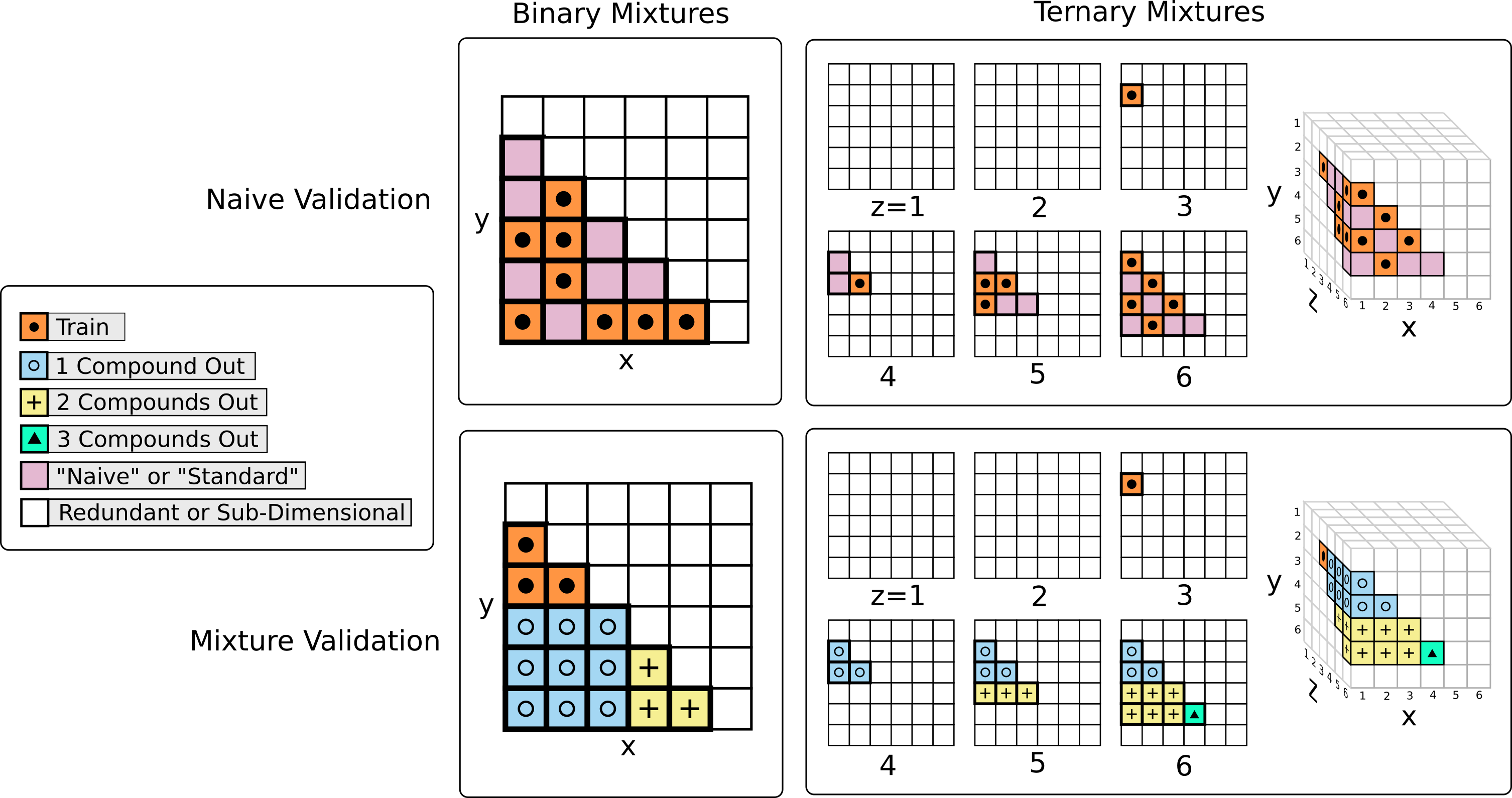}
 \caption{A visual comparison between standard and mixture validation on small binary and ternary datasets. The colors of the squares designate how they would be used in a modeling task. Orange squares are always the mixtures used for training, and the other colors are labeled with their relationship to the training set. For example, in the standard validation case, training and test mixtures are chosen randomly, which results in a random scattering of orange and purple squares. The ternary mixture visualization is shown as both a cube (right), and slices of that same cube (left) to show the squares that are occluded in the cube visualization. Note: white squares are either redundant or show mixtures that are below the dimension of the visualization (e.g., binary mixtures in the ternary case).
 }

 \label{fig:fulldiagram}
\end{figure}

Continuing with our example, we will detail the modifications needed to deal with ternary mixtures. Here there are now $3$ validation sets, corresponding to a degree of relatedness with the training set---or number of constituents drawn from the interior set---of $0$, $1$, and $2$. As in the binary example, we refer to these by the number of compounds drawn from the exterior set: `3 compounds out', `2 compounds out', and `1 compound out', respectively. This and the previous example are shown in Figure~\ref{fig:fulldiagram}.

The extension to arbitrary $N$-ary mixtures should now be clear. We will have $N$ validation sets, one for each of the possible numbers of `compounds out' from $1$ to $N$, in addition to the possibility of performing standard validation when applicable. The extension to other folds should also be clear: the procedure is the same but for a replacement of the interior and exterior chemical sets.

It should be strongly noted here that while we generally look disparagingly on the use of standard validation to mixtures, it is not always inappropriate. Specifically, when the intended use of a model is to fill in missing property entries in a set of mixtures, then standard validation most precisely mimics this outcome and is the preferred form of validation. In the language of the previous paragraph, this is the scenario with a maximal degree of relatedness between the training and validation datasets. In other words, the collection of mixture constituents is shared between the datasets.

\subsection{Estimating heritability with mixture pseudodescriptors}

Once a modeler has decided on an intended use or uses of a model and created the appropriate validation sets, we contend that validation is not complete without an estimate of the correlation between mixture properties in the training and validation datasets. A high correlation does not nullify the utility of a model; on the contrary, it can be an important part of the model. Instead, we argue that this correlation should be thought of as a baseline against which models are to be compared.

To determine heritability, modelers should repeat their modeling algorithm, using the same training and validation splits, but using mixture-specific pseudodescriptors instead of whichever descriptors were originally chosen. Like pseudodescriptors for single-chemical modeling~\cite{rucker_y-randomization_2007}, mixture-specific pseudodescriptors are random numbers or vectors; however, these random vectors are not assigned to each mixture separately but to each mixture constituent, and mixture descriptors are subsequently derived from the constituent pseudodescriptors. For example, the descriptor for a mixture may simply be the concatenation of each of its constituents' descriptors. This ensures that related mixtures have correlated descriptors while removing any structural information from the mixture descriptors. Models built with such descriptors will be unable to ascertain structure-property relationships and, therefore, any performance better than chance is owed to mixture heritability.

\subsection{Wide applicability of the mixture concept}

While we have outlined the validation sets for a simple example of $N$-ary chemical mixtures, there are many more datasets that can be thought of as mixture datasets, and therefore the validation strategies we describe are valid broadly with only minor modifications. Any dataset where the primary object of interest is itself comprised of multiple, more fundamental objects can be construed as a mixture dataset. For example, when modeling protein-ligand binding affinity, as in~\cite{jones_improved_2021}, for multiple protein targets and ligands we should view each protein-ligand complex as a mixture and validate according to the intended use of the model. In this case, the possible intended uses of the model are more nuanced than previously. Since the protein-ligand mixtures are now created by combinations from two distinct sets---the proteins and the ligands---rather than a single set of compounds as previously considered, we can imagine separately stratifying by the number of `proteins out' and `ligands out' rather than just by the number of `compounds out'.

In the above example, the notion of `proteins out' might most literally be taken to mean stratifying by exact protein sequence matches. However, small sequence modifications may not dramatically change the binding pocket, and we might therefore view two very similar protein sequences as actually the same for modeling purposes. In fact, we may cluster the proteins in a dataset according to sequence similarity and consider each cluster as a mixture constituent rather than each individual sequence.

Another example comes from~\cite{preuer_deepsynergy_2018}, where ternary mixtures of a single cell line and two drugs were considered. Again, we can distinguish between `cell lines out' and `compounds out'. Another ternary mixture example comes from DNA encoded libraries (DELs), which are libraries of single molecules often synthesized from three distinct building-block fragments. In~\cite{mccloskey_machine_2020}, DEL compounds, which can be considered for modeling purposes as ternary fragment mixtures, were modeled as binary mixtures to aid analysis. Finally, we claim that even simple mixture datasets supplemented with concentration data can be handled similarly, where now we view each concentration as itself a constituent of the mixture, and we can validate according to `concentrations out' in addition to `compounds out', etc.  For more detail on how concentrations can be construed as mixture constituents, consult appendix~\ref{sec:mixtures}.

\begin{table}[h!]
\begin{center}
\begin{tabular}{|c|c|}
\hline
\textbf{Mixture Dataset} & \textbf{Example References}  \\
 \hline
Drug-drug mixtures without concentration & ~\cite{noauthor_preparation_nodate,muratov_existing_2012} \\
\hline
Drug-drug mixtures with concentration & ~\cite{altenburger_mixture_2003,lu_joint_2009,lin_quantification_2003} \\
\hline
Protein-ligand binding & ~\cite{jones_improved_2021}\\
\hline
Drug-drug-cell line & \cite{preuer_deepsynergy_2018} \\
\hline
Fragment-fragment-fragment (DELs) & ~\cite{mccloskey_machine_2020-1}\\
\hline
\end{tabular}
\caption{Example mixture datasets with selected references demonstrating the wide applicability of the mixture concept.}
\label{tab:datasets}

\end{center}
\end{table}

\section{Examples: Real and Simulated Data}\label{sec:examples}
\subsection{Pancreatic cancer mixture synergy}\label{sec:ncatspaper}
In collaboration with researchers at NCATS and MIT, we have performed QSAR modeling on mixtures of drugs known to inhibit growth of pancreatic cancer \textit{in vitro} with the purpose of discovering binary combinations that exhibit synergy~\cite{noauthor_preparation_nodate}. The complete labeled dataset consisted of binary mixtures of $32$ different drugs, making $496$ mixtures, each labeled with a synergy score, namely a $\gamma$ score, from which a binary synergy label can be derived by means of a threshold~\cite{cokol_systematic_2011}. As part of our modeling effort, we investigated the performance of models in all three validation scenarios: standard, `1 compound out', and `2 compounds out'. It should be noted that since our data matrix was complete, i.e.\ there were no missing entries, there would be no purpose for standard validation, and it truly represents an irrelevant validation in this example. More precisely, since the synergy score is known for all binary mixtures formed with the $32$ drugs in our study, the only reasonable use of a model is to find new synergistic mixtures in either the `1 compound out' or `2 compounds out' category. We report in table~\ref{tab:performance} the performance of a single random forest model built using the concatenation of the sum and absolute difference of the constituent Morgan fingerprints as mixture descriptors. Specifically, we are reporting the area under the receiver operating characteristic curve (AUC-ROC) when using the real descriptors, mixture pseudodescriptors, and $y$-randomized labels with the real descriptors for each of the three validation strategies. As a reminder, our mixture pseudodescriptors are built by first randomizing the Morgan fingerprint of each constituent and then combining them in the same way the real mixture descriptors are created from their constituent fingerprints.

\begin{table}[h!]
\begin{center}
\begin{tabular}{|c|c|c|c|}
\hline
 & Standard & 1 Compound Out & 2 Compounds Out \\
 \hline
Morgan Fingerprints & $0.82\pm 0.01$ & $0.78 \pm 0.06$ & $0.66 \pm 0.17$ \\
\hline
Mixture Pseudodescriptors & $0.81 \pm 0.02$ & $0.72 \pm 0.08$ & $0.56 \pm 0.14$ \\
\hline
y Randomization & $0.51 \pm 0.05$ & $0.48 \pm 0.06$ & $0.44 \pm 0.20$ \\
\hline
\end{tabular}
\caption{Model performance on NCATS mixture data in each of three validation scenarios. Shown are the average and standard deviation of the area under the ROC curves for each of $5$ external folds. Actual fingerprints and mixture pseudodescriptors perform comparably in the standard and `compounds out' validation demonstrating a high level of correlation in the dataset.}
\label{tab:performance}

\end{center}
\end{table}

We can see from the table that mixture pseudodescriptors perform comparably to real features, especially in the standard validation setting. This demonstrates a high level of correlation between the training and validation datasets. We interpret from this that much of the model's performance is simply due to memorization of which compounds (independent of their chemical structure) are frequently found in synergistic or antagonistic mixtures. 

Compared to standard validation, we expect and observe lower model performance in the `1 compound out' validation where less memorization is possible due to the presence of novel mixture constituents. And while model performance with Morgan fingerprints does not deviate significantly from that of models built on mixture pseudodescriptors, this does not imply model failure. Instead, as in the standard case, it simply means that the models are relying heavily on the correlation (heritability) between mixture components and relying less heavily on the chemical structures of the components.

This statement no longer holds in the `2 compounds out' validation, as there can be no memorization, and thus no reliance on heritability, when training and validation mixtures share no individual compounds. Thus, we expect that modeling with mixture pseudodescriptors will produce an AUC-ROC of $0.5$ in this case, which is consistent with the data. Any enhancement in model performance from the use of Morgan fingerprints over mixture pseudodescriptors is an argument for the succesful incorporation of chemical structure into the model's predictions. This is because mixture pseudodescriptors are specifically built to lack constituent chemical structure.

\subsection{Simulated data example}

Here we give a toy example of ternary mixture data with the specific goal of demonstrating once again how standard validation can lead to a disastrous overestimate of a model's performance if the model were to be used in any of the various `$m$ compounds out' scenarios. We will consider ternary mixtures formed from a collection of $D$ drugs, labeled $d_i$ for $i = 1, \ldots, D$. Each drug is associated with a number $\alpha(d_i)$, representative of, for example, a chemical property. In this toy model, we will explicitly violate the foundational structure-activity hypothesis of SAR modeling by assuming that $\alpha(d_i)$ is randomly drawn from a standard Gaussian distribution,
\be
\alpha(d_i) \sim \mathcal{N}(0,1).
\ee

Let $\beta(d_i,d_j,d_k)$ represent a property of the mixture of drugs $d_i$, $d_j$, and $d_k$. We will assume $\beta$ is a stochastic variable centered on the sum of the associated $\alpha$ values,
\be
\beta(d_i,d_j,d_k) = \alpha(d_i) + \alpha(d_j) + \alpha(d_k) + \eta,
\ee
where $\eta \sim \mathcal{N}(0,\epsilon)$ for some $\epsilon >0$. This is a simulation of an experiment in which $\beta$ is noisily measured for each ternary drug mixture. More realistically, we would also assign noise to each drug's $\alpha$ value to represent our uncertainty in its measurement, but this will neither be necessary for nor will its absence be destructive to our toy model.

We will refer to a mixture $(d_i,d_j,d_k)$ with $\beta(d_i,d_j,d_k) > 0$ as `active' and assign a binary value of $1$ to it, while those mixtures with $\beta \leq 0$ are `inactive'. On average, half the mixtures will be active, and so an accuracy of roughly half will be our baseline for assessing a model's worth.

We have performed this toy experiment with $D = 32$ drugs, resulting in $4960$ mixtures, a large enough number to yield consistent results. Each of the simulated drugs used in this model was assigned a random, $128$-dimensional binary fingerprint as a feature vector. We trained a random forest model to predict the activity class of mixtures according to both the naive and mixture validation schemes. Results are shown in table~\ref{tab:toyperformance}.

\begin{table}[h!]
\begin{center}
\begin{tabular}{|p{3.5cm}|c|c|c|c|}
\hline
 & Standard & 1 Cmpnd Out & 2 Cmpnds Out & 3 Cmpnds Out \\
\hline
Mixture \newline Pseudodescriptors & $0.80 \pm 0.04$ & $0.76 \pm 0.06$ & $0.69 \pm 0.10$ & $0.52 \pm 0.09$ \\
\hline
y Randomization & $0.53 \pm 0.02$ & $0.53 \pm 0.02$ & $0.52 \pm 0.05$ & $0.58 \pm 0.05$\\
\hline
\end{tabular}
\caption{Toy model performance on simulated mixture dataset. Shown here are the average and standard deviation of accuracy over $5$ fold cross-validation in each of the modeling/validation scenarios.}
\label{tab:toyperformance}

\end{center}
\end{table}

First, as would be expected, the model performed no better than the baseline set by the $y$-randomized model, itself consistent with $50\%$ accuracy, when using the `3 compounds out' in this case, validation set. This is expected because there is no structure-activity relationship for the model to learn. At best, the model can learn $\alpha(d_i)$ for each of the drugs in the training set, leaving it completely ignorant of the $\alpha$ property of those drugs not seen by the model.

We expect increasing performance of the model with increasing relatedness between the training and validation sets, and we find this to be true. This reflects the model's ability to learn $\alpha(d_i)$ for those drugs in the training set. Again, it should be stressed that this learning is purely memorization and reflects no generalizable knowledge, as there is none to be had in this toy model. Alarmingly, this memorization can yield a performance of $80\%$ accuracy in the standard validation scenario. All experiments were performed with $\epsilon^2 = 0.5$. Recall that $\epsilon$ controls the noise of the $\beta$ measurement. With smaller values of $\epsilon$, the model's supposed performance can be even higher.

\section{Conclusions}
In summary, we have demonstrated a common pitfall when validating models built on mixture data, and we have suggested methods to avoid it. In particular, we have noted that mixture datasets are rife with statistical correlations not found in single-chemical datasets, and---if not properly accounted for---these correlations can lead modelers astray as to the expected performance of their models. To account for these correlations, we propose that modelers must first identify the intended purpose of their model, stratified by the number of mixture constituents not present in the training data at the time of model testing. For each intended purpose, modelers must then ascertain the baseline level of correlation in the data by performing modeling with mixture pseudodescriptors as outlined in the text. Model performance above the level of the pseudodescriptor baseline is indicative of a model’s incorporation of structural features into its predictions.

\appendix

\section{$N$-ary Mixtures}\label{sec:mixtures}
\subsection{The structure of mixture data}
In the following, we will establish a working definition of a mixture as well as useful notation for dealing with mixtures. The latter will be inspired by the mathematics of sets. In particular, a mixture is an element of the Cartesian product of a collection of sets defining the types of objects making up the mixture. For example, we may be interested in binary mixtures of drugs, one drawn from each of two different collections. Labeling the two collections of drugs as $\mathcal{A}$ and $\mathcal{B}$, respectively, then a mixture $m$ is an element of $\mathcal{A}\times \mathcal{B}$, denoted\footnote{The symbol $\in$ means that what is on the left is an element of the set on the right}
\be
m =(a,b) \in \mathcal{A}\times \mathcal{B},
\ee
where $a \in \mathcal{A}$ and $b \in \mathcal{B}$. In words, the separate drugs $a$ and $b$ comprise the mixture $m$, with drug $a$ drawn from collection $\mathcal{A}$ and drug $b$ drawn from collection $\mathcal{B}$. This definition easily incorporates mixtures made from more than two drugs by increasing the number of collections. For example, a three drug mixture is an element of $\mathcal{A}\times\mathcal{B}\times\mathcal{C}$:
\be
m = (a,b,c) \in \mathcal{A}\times\mathcal{B}\times\mathcal{C},
\ee
where $\mathcal{C}$ is now a third collection of drugs.

There are two immediate issues with this definition. The first centers around the issue of ordered vs.\ unordered mixtures. In the previous examples, we made the implicit assumption that each of the drug collections was distinct from the rest, i.e.\ that no drug belonged to more than one collection. In this case, a drug mixture such as $(a,b)$ is inherently ordered, as the first drug, $a$, is defined to come from collection $\mathcal{A}$, which is disjoint from collection $\mathcal{B}$ containing the second drug, $b$.

However, it is often of interest to study mixtures of drugs taken from the same collection or set, possibly consisting of all drug-like chemicals or those with known activity in an assay of interest. Let us denote this collection as $\mathcal{D}$. Then, a binary mixture of drugs from collection $\mathcal{D}$ consists of two drugs $(d_1,d_2)$, but---despite the numerical naming---there is no order associated with such a mixture. The mixture is just as well written $(d_2,d_1)$.

Mathematically, we cannot treat this mixture as element of the Cartesian product $\mathcal{D}\times \mathcal{D}$, as within this set the two orderings---$(d_1,d_2)$ and $(d_2,d_1)$---are considered distinct. The correct notion is that of a quotient set or equivalence class, but we will be satisfied with thinking in terms of Cartesian products and informally keeping track of when ordering does or does not matter.

The next issue with our definition of a mixture is that it seemingly fails to incorporate relative concentrations when applicable. Notably, relative concentrations are not always applicable, such as when the relevant outcome being measured is not a concentration-dependent quantity but instead a summary of many concentration-dependent quantities. This is the case for various measures of synergy, which depend on the deviation of an entire mixture dose-response surface rather than individual points on the surface. We will address this point in more detail later, when we describe the broad validity of this notion of mixture, but it will suffice for now to say that inclusion of relative concentrations is possible by treating the relative concentrations of the mixture as separate constituents of a broader notion of mixture.

Finally, we conclude with the structure of mixture data and illuminating examples. A dataset of $N$-ary mixtures may, on the one hand, be considered as simply a collection of mixtures $m \in \mathcal{M}$; however, the mixture structure is better appreciated when each of the constituent collections is elucidated,
\be
m = (a_1, a_2, \ldots, a_N) \in \mathcal{A}_1\times\mathcal{A}_2\times \ldots \times \mathcal{A}_N \supset \mathcal{M}.
\ee
Note that we have specified that the Cartesian product of constituent collection contains as a set the collection of mixtures in the dataset, $\mathcal{M}$. This occurs when not all possible mixtures are present in the data. This will be the case in many scenarios but it will complicate the exposition of validation strategies detailed below. So, we will assume instead that $\mathcal{A}_1\times\mathcal{A}_2\times \ldots \times \mathcal{A}_N = \mathcal{M}$.

Visually, a collection of binary mixtures can be represented by a rectangular array (or matrix), with each entry in the array representing a specific mixture. Missing entries correspond to specific mixtures not present in the dataset, a common occurrence---as mentioned previously---but not one we will deal with in detail, as the modifications to our proposed validation strategies are straightforward but unsightly. Therefore, we will only consider complete mixture matrices with all entries filled. The extension of this visual to $N$-ary mixtures, while more difficult to commit to paper, is conceptually simple. Here, the collection of $N$-ary mixtures comprise an $N$-dimensional, rectangular array. For examples of ternary mixture collections, see Figure~\ref{fig:fulldiagram}.

We have already described mixtures of drugs, both when each drug comes from a distinct collection and when all drugs are drawn from the same collection. While this may be the most intuitive notion of a mixture, this concept extends much more broadly and applies to many datasets. Examples include~\cite{preuer_deepsynergy_2018}, wherein a mixture consists of two drugs and a cell line used in an assay. This is an example of a ternary mixture. Another ternary mixture example comes from DNA encoded libraries (DELs), which are libraries of single molecules often synthesized as a mixture of three distinct fragments. In~\cite{mccloskey_machine_2020}, ternary fragment mixtures from DELs were converted into binary mixtures to aid analysis. Continuing, an unconventional binary mixture example comes from attempts to model protein-ligand binding affinity, such as in~\cite{jones_improved_2021}. Here, a mixture is a protein-ligand complex.

Lastly, we have already mentioned that a dataset of chemical mixtures including relative concentrations can be construed as a broader, mixture dataset when the relative concentrations are themselves included as constituents. This point can be made clearer when we imagine binning or discretizing the allowed concentrations, thus making for a finite collection of possible concentrations to be included. As a specific example, let us consider binary mixtures composed of one drug drawn from collection $\mathcal{A}$ and one drug drawn from collection $\mathcal{B}$ mixed together at a relative concentration from the collection $\mathcal{C}$. Then, our complete collection of `mixtures' (including concentrations) is the set
\be
\mathcal{M} = \mathcal{A} \times \mathcal{B} \times \mathcal{C}.
\ee
The reader may already recognize this as identical to our example before of a general ternary mixture, and we claim it is no different.

In each of these examples---and any other modeling scenario for which the dataset is conceivably considered as a mixture dataset---care must be taken when choosing validation strategies, which we will detail presently.

\subsection{Standard validation}
As discussed previously, a validation strategy includes the process of decomposing a dataset into separate training and validation/testing subsets. Using the relationship
\be
\mathcal{M} = \mathcal{A}_1 \times \mathcal{A}_2 \times \ldots \times \mathcal{A}_N,
\ee
one can decompose a dataset of mixtures either by considering the mixtures as independent entities or by separately decomposing each of the constituent subclasses and subsequently forming mixtures. This distinction will be the essential difference between what we call `standard' and `mixture' validation. 

Absent a mixture-specific representation or descriptor, most modeling for mixtures will be ultimately based on descriptors of the mixture constituents\footnote{While there do exist some mixture-specific descriptors, most notably \cite{kuzmin_simplex_2021}, these still incorporate descriptors for each constituent separately as a portion of the full mixture descriptor. Therefore, the statistical dependence we refer to still applies.}. Therefore, a rational validation strategy should also be based on the constituents. This is because validation sets are assumed to be uncorrelated with training sets, which is a necessary assumption for external validity of the results.\footnote{In some cases---discussed more in the introduction and later---training and validation sets being uncorrelated is too strong a requirement. However, in all cases, the level and type of correlation needs to be considered.} However, if training sets and validation sets have overlapping constituents, there is good reason to believe these datasets are correlated in ways that can obscure a model's real performance.

Standard validation is therefore the process of decomposing the set of mixtures randomly into training and validation sets without consideration for the correlation between these sets. There are only limited circumstances where the validation performance of a model in this scenario accurately reflects the performance of the model in its intended use.

\subsection{Mixture validation}
To explain the various forms of mixture validation we propose, it will be easiest to consider the most basic of mixture datasets: each mixture is a collection of $N$ drugs drawn from the same collection, which we label $\mathcal{D}$.\footnote{To connect this to our previous definition, we have $\mathcal{M} = \mathcal{D}^N = \mathcal{D} \times \mathcal{D} \times \ldots \times \mathcal{D}$, informally. More precisely, this is a scenario where ordering does not matter, and we need to keep that in mind.} Let us suppose that $\mathcal{D}$ contains $D$ total drugs, and we can label them---randomly---as $d_1$, $d_2$, etc., up to $d_D$. As described in the body, best practices in any validation, mixture model or otherwise, calls for $k$-fold cross-validation. Often $k$ is chosen equal to $5$, but we will leave it arbitrary for completeness. For simplicity, we will assume that $k$ divides $D$, so that $D = kk'$ for some integer $k'$. This isn't strictly necessary, and the required modifications of our proposal are straightforward. However, this assumption makes the explanation cleaner.

Our simplifying assumption allows us to split the collection of drugs into $k$ sets each of size $k'$. Without any loss of generality, we will split them just by their order, so that the first set includes drugs $d_1$ through $d_{k'}$, and so on. Recall that $k$-fold cross-validation calls for $k$ pairs of training and validation sets, with all of the training sets commonly taken to be mutually disjoint. Again without loss of generality, we will describe the first of the $k$-folds with the remaining folds easily inferred.

The first training set consists simply of those mixtures derived solely from drugs in the set $\{d_1, \ldots, d_{k'}\}$. We will call this the interior set. Note that we don't refer to it as the training set, because the interior set consists of drugs and not mixtures, whereas the training is actually performed on mixtures. Additionally, we will refer to the remaining drugs, $\{d_{k'+1}, \ldots, d_{D} \}$, as the exterior set.

If we are modeling $N$-ary, order-independent mixtures, the training set will have size $k' \choose N$. Unlike single-chemical modeling, however, we now have $N$ validation sets, each relevant for a different intended use of the model. Let us refer to these sets as $\mathrm{Val}_m$, where $m = 1, \ldots, N$. $\mathrm{Val}_m$ is defined as containing only those mixtures that include exactly $m$ drugs in the exterior set. Equivalently, they are the mixtures including exactly $N - m$ drugs in the interior set. In the terminology of the rest of this paper, $\mathrm{Val}_m$ is referred to as `$m$ compounds out'.

We note that with this definition $\mathrm{Val}_0$ is the training set. Furthermore, $\mathrm{Val}_N$ contains no drugs from the internal set. In~\cite{muratov_jsm_2014}, this is the set referred to as `everything out', while we prefer the more generalizable `$N$ compounds out'. In the binary mixture case, $N =2$, the only remaining set is $\mathrm{Val}_1$, previously referred to as `compounds out' and here called `1 compound out'. For $N>2$, all of the sets $\mathrm{Val}_m$ for $m\neq N$ are some generalization of `compounds out' validation sets.

Let us briefly comment on the modifications needed to deal with the more general case where
\be
\mathcal{M} = \mathcal{A}_1 \times \mathcal{A}_2 \times \ldots \times \mathcal{A}_N.
\ee
Let us only focus on describing a single split or fold, with the remaining folds defined similarly. Each of the constituent sets $\mathcal{A}_i$ is split into an `interior' and an `exterior' set, denoted respectively as $\mathcal{I}_i$ and $\mathcal{E}_i$, where $i =1 ,\ldots,N$ stands in for any the $N$ constituent collections. We require that the interior and exterior sets are disjoint,
\be
\mathcal{I}_i \cap \mathcal{E}_i = \varnothing, \quad \forall i =1 ,\ldots, N.
\ee
The training set $\mathcal{T}$ is the product of all interior sets
\be
\mathcal{M} \supset \mathcal{T} = \mathcal{I}_1 \times \mathcal{I}_2 \times \ldots \times \mathcal{I}_N \subset \mathcal{A}_1 \times \mathcal{A}_2 \times \ldots \times \mathcal{A}_N.
\ee
As for validation sets, now there are many more than in the previous, simple case. To illustrate, let us consider what is now the analogue of $\mathrm{Val}_{N-1}$. In the previous case, recall that this set includes all mixtures composed of $N-1$ drugs from the exterior set and $1$ drug from the interior set. Now, with $N$ interior and exterior sets, there are extra distinctions to be made. Specifically, now it must be decided which of the ordered $N$ constituents is to come from the interior set. As an example, there is a difference between the sets
\be
\mathcal{I}_1 \times \mathcal{E}_2 \times \mathcal{E}_3 \ldots \mathcal{E}_N \neq \mathcal{E}_1 \times \mathcal{I}_2 \times \mathcal{E}_3 \ldots \mathcal{E}_N.
\ee
In this case, what was $\mathrm{Val}_{N-1}$ has fractured into $N$ distinct versions, one for each of $N$ choices. Generally, $\mathrm{Val}_m$ will fracture into $N \choose m$ distinct validation sets. In total, there are $2^N -1$ validation sets.

\section*{Acknowledgements}
During the course of this work, T.M.\ was supported by the National Institute of General Medical Sciences of the NIH under Award Number T32GM086330. J.W. is supported but the National Institute of General Medical Sciences of the NIH under Award Number T32GM135122

\newpage
\printbibliography

\end{document}